\documentstyle[aps,prb,twocolumn]{revtex}
\begin{document}
\draft
\title{Quantum states in a magnetic anti-dot}
\author{J. Reijniers and F. M. Peeters \cite{peeters}}
\address{Departement Natuurkunde, Universiteit Antwerpen (UIA), \\
Universiteitsplein 1, B-2610 Antwerpen, Belgium.}
\author{A. Matulis \cite{matulis}}
\address{Semiconductor Physics Institute,\\ 
Go\v{s}tauto 11, 2600 Vilnius, Lithuania. }
\date{\today}
\maketitle
\begin{abstract} 
We study a new system in which electrons in two dimensions are confined by a non homogeneous magnetic field.  The system consists of a heterostructure with on top of it a superconducting disk.  We show that in this system electrons can be confined into a dot region.  This magnetic anti-dot has the interesting property that the filling of the dot is a discrete function of the magnetic field.  The circulating electron current inside and outside the anti-dot can be in opposite direction for certain bound states.  And those states exhibit a diamagnetic to paramagnetic transition with increasing magnetic field.  The absorption spectrum consists of many peaks, some of which violate Kohn's theorem, and which is due to the coupling of the center of mass motion with the other degrees of freedom. 
\end{abstract}


\section{Introduction}
Quantum dots have been the subject of both theoretical and experimental research in recent years.\cite{johnson}  They have been successfully created experimentally by applying lithographic and etching techniques to impose a lateral structure onto an otherwise two-dimensional electron system.  These structures introduce electrostatic potentials in the plane of the two-dimensional electron gas (2DEG), which confine the electrons to a dot region.  The energy levels of electrons in such a quantum dot are fully quantized like in an atom, and therefore are also referred to as \emph{artificial atoms}.  In such electrically confined quantum dots the confinement potential can well be represented by a parabolic potential.

In the present paper we study a new quantum dot system which is different from the usual quantum dot system: 1) the electrons are confined magnetically, 2) the confinement potential is inherently non-parabolic, and 3) the dot contains a finite number of electrons where the filling of the dot is a discrete function of the strength of the confinement (magnetic field).

This \emph{magnetic anti-dot} can be realized \cite{geim97} by depositing a superconducting disk on top of a 2DEG.  When a homogeneous magnetic field is applied perpendicular to the 2DEG, the magnetic flux lines are expelled from the superconducting disk due to the Meissner effect, which results in an inhomogeneous magnetic field profile in the 2DEG.  Note that this problem is related to the problem of type II superconductors on a heterostructure, where flux lines penetrate the 2DEG.\cite{brey93}  Between the flux lines regions of zero magnetic field are present.  Here we have the inverse situation in which we have a uniform magnetic field except for a local dot-like region where there is no magnetic field present.  Bound states are now possible in the zero magnetic field region.

Such a system was discussed by Peeters \emph{et al.} in Ref.~\onlinecite{peeters96} where preliminary results were presented for the energy levels of such a magnetic anti-dot.  Here we elaborate on this system and give a more detailed and complete study of the bound states of such a system and calculate also other properties of this system.  In Ref.~\onlinecite{peeters98} the Hall and bend resistance resulting from such a system was discussed in the ballistic regime and in Ref.~\onlinecite {ibrahim98} the diffusive transport of such a magnetic anti-dot placed on top of a Hall bar was studied. 

Solimany \emph{et al.}\cite{solimany95} studied a limiting case of the present system in which the magnetic field was $B(\rho)=B_a \delta(\rho-a)$ and they solved the classical and quantum mechanical equations for a magnetically confined quantum dot, and recently Sim \emph{et al.}\cite{sim98} investigated the formation of magnetic edge states along with the corresponding classical trajectories.  Although both groups referred to this theoretical system as a magnetic dot, we think the name `magnetic anti-dot' is more appropriate in this case.  Here we start from the experimental realizable system, i.e. a superconducting disk on top of a 2DEG, and calculate in Sec.~II the non-homogeneous magnetic field profile induced in the 2DEG.  We find that the magnetic field profile is different from the one assumed in Refs.~\onlinecite{solimany95} and \onlinecite{sim98}. In Sec.~III the energy spectrum of electrons in the 2DEG near this magnetic field profile is calculated and compared with those of two circular model magnetic anti-dot systems.  The filling of the magnetic anti-dot is calculated as function of the strength of the confinement.  In Sec.~IV we discuss the electron probability current and the induced magnetic moment.  The optical absorption spectrum, i.e. frequencies and oscillator strengths, are obtained in Sec.~V.  The results are summarized and our conclusions are presented in Sec.~VI.

\section{Magnetic field profile}
The system we have in mind is shown schematically in the inset of Fig.~\ref{fig:profile}.  We have a high mobility heterostructure with on top of it a superconducting disk placed in a homogeneous applied magnetic field $\textbf{B}=(0,0,B_a)$.  For convenience, we consider a very thin superconducting disk which is perpendicular to the magnetic field and a distance $z$-above the 2DEG.  Because of the symmetry of the system, we use cylindrical coordinates ($\rho,\,\varphi,\,z$) and measure all lengths in units of the disk radius $a$.   We solve the following magnetostatic equations outside the disk

\begin{equation}
\begin{array}{lll}
  \mathbf{\nabla} \cdot \mathbf{B} &=& 0, \nonumber \\
  \mathbf{\nabla}\times \mathbf{B} &=& 0,
\end{array}
\end{equation}
with the conditions that $B_{\varphi}= 0$ due to symmetry considerations, and $B_z|_{\rho\le1} = 0$ on the disk surface (z=0).

This problem is equivalent to the problem of liquid flow around a disk and was solved in Ref.~\onlinecite{morse53}.  There the \emph{oblate spheroidal coordinates} $\rho = \sqrt{(\xi^2+1)(1-\eta^2)}$, $\varphi$ and $z=\xi\eta$ were introduced, where $0<\xi < \infty$, $0<\eta<1$ and $0<\varphi<2\pi$ which have the corresponding scale factors $h_{\xi}=\sqrt{(\xi^2+\eta^2)/(\xi^2+1})$, $h_{\eta}=\sqrt{(\xi^2+\eta^2)/(1-\eta^2})$, $h_{\varphi}=\sqrt{(\xi^2+1)(1-\eta^2)}$. The surface $\xi = 0$ is a disk of radius 1 in the $xy$-plane with center at the origin.  According to that solution, the magnetic field under such an infinitesimal thin disk can be represented as ${\mathbf B}=-{\mathbf \nabla}  \Phi$ where 
\begin{equation}
  \Phi(\xi, \eta) = -B_a \eta \left\{\xi + \frac{2}{\pi}\left[ 1-\xi \tan^{-1}
  \left(\frac{1}{\xi}\right)\right]  \right\}.
\label{eq:potential}
\end{equation}
Using cylindric symmetry and $\mathbf{B}=\nabla \times \mathbf{A}$, we arrive at the single non zero component of the vector potential
\begin{equation}
  A_{\varphi}(\xi,\eta) = \frac{B_a\rho}{2}
  \left\{1 +\frac{2}{\pi}\left[\frac{\xi}{1+{\xi}^2} - \tan^{-1}
    \left(\frac{1}{\xi}\right)\right] \right\},
\end{equation}
from which we obtain the perpendicular component of the magnetic field profile (or from Eq. (\ref{eq:potential}) using ${\mathbf B}=-{\mathbf \nabla}  \Phi$) 
\begin{equation}
B_z(\xi, \eta)=\frac{2A_\varphi}{\rho}+\frac{2B_a}{\pi}\frac{\xi(1-\eta^2)}{(1+\xi^2)(\xi^2+\eta^2)}.
\end{equation}
These results can be easily converted back to the cylindrical coordinate system ($\rho,\,\varphi,\, z$) by inserting $\xi^2=(1/2)[\sqrt{(r^2-1)^2+4z^2}+(r^2-1)]$ and $\eta^2=(1/2)[\sqrt{(r^2-1)^2+4z^2}-(r^2-1)]$ with $r^2=\rho^2+z^2$ in the above equations.

The resulting magnetic field profile is shown in Fig.~\ref{fig:profile}(a) for different values of the distance ($z$) between the 2DEG and the superconducting disk.  Notice that the magnetic field under the disk is very small due to the Meissner effect, while far from the disk it becomes equal to the external magnetic field strength $B_a$.  At the edge of the disk there is an overshoot of the magnetic field strength, which becomes larger with decreasing value of $z$.  When the 2DEG is further away from the superconducting disk, there is almost no overshoot and $B_z$ gradually increases away from the center.  Notice that in the latter case the magnetic field is nonzero and consequently the model systems of Ref~.\onlinecite{solimany95} and Ref.~\onlinecite{sim98} can not be realized by using a thin superconducting disk.

Before discussing the electron states in such a magnetic field profile, we will first consider two model systems which correspond to two extreme situations, but which contain the essential physics of the problem.  The two models we consider are defined by the following profiles: 1) $B_z(\rho)=B_a \theta(\rho-1)$ with corresponding vector potential $A_\varphi(\rho)=[B_a(\rho-1/\rho)/2]\theta(\rho-1)$, and 2) $A_\varphi=(B_a\rho/2)\theta(\rho-1)$, which results into a magnetic field profile with a delta function overshoot $B_z=B_a\theta(\rho-1)+(B_a\rho/2)\delta(\rho-1)$. The first model was used as the magnetic field profile in the magnetic anti-dot of the previous papers \cite{peeters96,peeters98,ibrahim98,solimany95,sim98}.  The second model contains a magnetic overshoot at the edge of the magnetic dot (see also Ref.~\onlinecite{peeters96}), which models the real system for $z/a\ll 1$.  This overshoot is due to the de-magnetization effects of the superconducting disk.  The vector potential profile of these two models is shown in Fig.~\ref{fig:profile}(b) together with the one resulting from the superconducting disk for two different values of the set back distance $z$.     

\section{The energy spectrum}
In order to calculate the wavefunctions and their corresponding energy, we have to insert the expression for the spatial dependent vector potential ($\mathbf{A}$) into the momentum operator ${\mathbf p} \rightarrow {\mathbf p}+ (e/c) {\mathbf A}$ which results in the Schr\"{o}dinger equation: $-(1/2m_e)[\hbar \Delta+(ie/c){\mathbf A}({\mathbf r})]^2 \Psi({\mathbf r})=E \Psi({\mathbf r})$.  Due to the cylindrical symmetry of the problem, the wave function can be presented as 
\begin{equation}
\Psi(\rho,\varphi)=\frac{1}{\sqrt{2\pi}}e^{im\varphi}R(\rho),
\end{equation}
and the problem reduces to solving the 1D radial equation
\begin{equation}
\left\{\frac{d^2}{d\rho^2}+\frac{1}{\rho}\frac{d}{d\rho}+2[E-V(\rho)]\right\}R(\rho)=0,
\end{equation}
where the effective potential reads
\begin{equation}
V(\rho)=\frac{1}{2}\left(A_\varphi(\rho)+\frac{m}{\rho}\right)^2.
\end{equation}

\noindent
We solved this eigenvalue problem numerically, using the Newton iteration technique and subjecting the solution to the following boundary conditions: $R(\rho\rightarrow 0)=\rho^{|m|}$ and $R(\rho\rightarrow \infty )=0$.

The numerical results for the energy spectrum are shown in Fig.~\ref{fig:spectra1}(a) for the model without overshoot, in Fig.~\ref{fig:spectra1}(b) for the model with overshoot and in Fig~\ref{fig:spectra1}(c) for the real profile in case of $z/a=0.01$.  The different energy levels are labeled with the corresponding quantum numbers ($n,m$).  We found it convenient to express the energy in units of $E_0=\hbar ^2/m_ea^2$ and the applied magnetic field $B_a$ in units of $B_0=c\hbar /ea^2$.  These units are related to the problem of a particle in a box.  For example for $a=0.01,\, 0.1,\, 1,\, 10\mu$m we have respectively $E_0=7.63 \times 10^{-1}, \, 7.63 \times 10^{-3},\, 7.63 \times 10^{-5},\, 7.63 \times 10^{-7}$ meV and $B_0 = 6.6\times 10^{4}, \, 6.6 \times 10^{2},\, 6.6,\, 0.066$ Gauss.  From Fig.~\ref{fig:spectra1} we notice that for small magnetic fields the energy is linear in $B_a$ and in fact we recover the Landau levels $E_{n,m}=\hbar \omega_c (n+(|m|+m)/2+1/2)$ for an electron in a homogeneous magnetic field.  The reason is that for small magnetic field we have $l_B/a \gg 1$, where $l_B= \sqrt{\hbar c/e B}$ is the magnetic length.  So the electron wavefunction is spread out over a large region and most of its probability is in the $\rho/a > 1$ region where $B(\rho) \approx B_a$.  In the opposite case of a large external magnetic field we have $l_B/a \ll 1$ and the electron wavefunction is localized in the dot region where there is no magnetic field present.  The problem is then similar to the one of an electron in a circular dot in the absence of a magnetic field and we recover the discrete energy levels which are determined by the zeros of the Bessel function: $J_{|m|}(k)=0$ and are plotted on the right hand side of the figure.   In this case the energy scale is $E_0=\hbar^2/m_ea^2$ like for a particle confined in a 1D box.

Although the limiting behavior of the spectrum for $B_a \rightarrow 0$ and $B_a \rightarrow \infty$ is very similar in all three cases the intermediate behavior turns out to be very different.  Indeed we see that the energy levels $E_{n,m}$ of the model without overshoot increase slower than linear with increasing $B_a$.  This is not the case when there is a magnetic field overshoot where there are a number of energy levels which for small/intermediate fields have a superlinear behavior as function of $B_a$.  This distinct behavior is made more visible when we plot the energy in units of $\hbar \omega_c$ as function of $1/l_B \sim \sqrt{B_a}$ as is done in Fig.~\ref{fig:spectra2}.  This different behavior between the two cases can be understood by looking at Fig.~\ref{fig:wavefunction}, where the radial part of the electron wavefunction (n,m)=(0,-1) is plotted  for various values of the magnetic field strength $B_a$ in case of the model with magnetic overshoot.  For $m<0$ the wavefunction exhibits a maximum at $\rho = \rho^{*}>0$.  With increasing $1/l_B$ this maximum shifts towards the center of the dot.  In case there is an overshoot in the magnetic field profile and when the maximum of the electron wavefunction is situated near $\rho/a=1$, the electron energy will be increased which results in the local maximum in the energy as shown in Fig.~\ref{fig:spectra2} .  From this interpretation it is easy to understand that when $m<0$ the maximum in $E_{n,m}/\hbar\omega_c$ shifts to larger $1/l_B$ with increasing $|m|$.

This behavior of the electron energy has important consequences for the filling of the dot.  Outside the quantum dot the magnetic field is $B_a$ and the electron lowest energy state is $\hbar\omega_c/2$.  An electron will only be situated in the dot when its energy is lower than in the region outside the dot.  From Fig.~\ref{fig:spectra2}(a)  we note that for a dot without magnetic field overshoot at its edges there are an infinite number of states, i.e. the states with $m \le 0$ for $n=0$, which have an energy less than $\hbar \omega _c/2 $ and consequently the electrons will be attracted towards the dot.  For the system with magnetic overshoot the situation is totally different.  The $|0,0 \rangle$ state has an energy below $\hbar \omega_c/2$ and two electrons (two because of spin) will be able to occupy the dot.  When we try to add more electrons to the system we see that for small $1/l_B$ the electrons will prefer to sit far away from the dot region where they have a lower energy, i.e. $\hbar \omega_c/2$.  Thus in this situation the electrons are repelled by the anti-dot.  Increasing the magnetic field will bring the (0,-1) level below $\hbar \omega_c$ and then two more electrons will be attracted towards the magnetic anti-dot.  This discrete filling of the dot is shown in Fig.~\ref{fig:filling} for the model with magnetic overshoot (solid curve) and for the superconducting disk case (dashed curve) with $z/a=0.01$. 

Including the real magnetic profile does not alter our conclusions qualitatively.  This is shown for $z/a=0.01$ in Fig.~\ref{fig:spectra1}(c) and Fig.~\ref{fig:spectra2}(c). We obtain some kind of intermediate behavior between the two model systems.  The discrete filling of the dot is still present, which can be inferred from Fig.~\ref{fig:spectra2}(c).  Nevertheless, the exact position at which the number of electrons jump to higher values is a function of the exact magnetic profile, and in particular depends strongly on the sharpness of the magnetic overshoot.  This is why $z/a$ has to be very small, which can always be experimentally achieved, by making the superconducting disk large enough.  

\section{The probability current distribution and the magnetic moment}
The probability current distribution of the eigenstates is also different from the usual quantum dot case.  When we rewrite the wavefunction as $\psi({\mathbf r})=\alpha({\mathbf r})e^{i\xi ({\mathbf r})}$, the probability current is given by \cite{cohen77}
\begin{equation}
{\mathbf J}({\mathbf r})=(1/m_e) \alpha^2 \left[\hbar \nabla \xi({\mathbf r})+(e/c) {\mathbf A}({\mathbf r}) \right],   
\end{equation}
where the first term is the well-known circular current and the second term is due to the magnetic field.  For bound states the current vector has only an angular component $\textbf{J}(\textbf{r})=J_{\varphi}(\textbf{r}) {\mathbf e}_{\varphi}$ which is independent of $\varphi$.  The probability current distribution $J_\varphi(\rho)$ and the corresponding radial distribution function of bounded states with $n=0$ and different $m$-values (indicated in the figure) are plotted in Fig.~\ref{fig:currentprofile1} for the three different cases.  For $m=0$ no current flows inside the dot except for the realistic magnetic field profile (Fig.~\ref{fig:currentprofile1}), although it is very small. The $m>0$ states have a positive circular current which is larger with increasing overshoot of the magnetic field at the edge.  For $m<0$ the current is circulating in the opposite direction and the magnitude increases with increasing overshoot.  The maximum of the current profile moves closer to the edge with increasing $|m|$.  Sufficiently outside the dot region the circular current is positive, irrespective of the value of $m$.  With the delta overshoot this is also true near the outside edge of the dot where the current distribution exhibits a discontinuous behavior.  In the latter case the current intensity is a uniform decreasing function of the distance $\rho > a$.  For the case without overshoot, or when we have a continuous overshoot, the current is negative near the outside edge for $m<0$.  These results can be understood from classical trajectories of magnetic edge states circulating clockwise ($m<0$) or counterclockwise ($m\ge 0$) along the boundary region of the magnetic anti-dot without overshoot (Cfr. Ref.~\onlinecite{sim98}) and which are shown in Fig.~\ref{fig:trajectories} for $m=-1$, $m=0$ and $m=1$.  For completion, we also included Fig.~\ref{fig:currentprofile2}, which is the same as Fig.~\ref{fig:currentprofile1}, but for the states with $m=1$ and different $n$-values (indicated on the figure). Now the radial part of the wavefunction has $n$-nodes which results in zeros in the circular current.  Note also that the current becomes much more strongly peaked near $\rho \approx 0$.  The amount of current outside the dot also increases with increasing $n$.  Because $m=1>0$ the current is positive irrespective of the value of $n$.

Also the magnetic moment is different from the quantum dot case. The magnetic moment of a particular bound state $|n,m\rangle $ is defined as follows: $M_{n,m} \equiv q/(2m_e)\langle n,m|\lambda_z|n,m \rangle$, where $\lambda_z$ is the $z$-component of the moment of the mechanical momentum {\boldmath $\lambda$}$={\mathbf r}\times[{\mathbf p}+(e/c){\mathbf A}]$.  For convenience we write the magnetic moment $M_{n,m}$ in units of $M_0=-e\hbar/(2m_e)$, so we obtain 
\begin{equation}
M_{n,m}= m+ \frac{e}{\hbar c}\langle n,m|\rho A_\varphi(\rho)|n,m\rangle,
\end{equation}
which is plotted for various one electron states in Fig.~\ref{fig:moment} for the three different systems.  In the limit $B_a\rightarrow 0$ we obtain the well known result $M_{n,m}=2n+|m|+m+1$: the magnetic moment is that of an electron in a homogeneous magnetic field.  Notice that all states with the same $n$ but $m \le 0$ have the same moment in this limit.  Furthermore $M_{n,m}>0$ for all bound states with $m\ge 0$.  For $B_a\rightarrow \infty$, the same result as for a circular dot defined by hard walls, i.e. $M_{n,m}=m$, is obtained for the models with (a) and without (b) overshoot.  Thus the magnetic moment of the $m<0$ states changes sign with increasing $B_a$.  This change in sign occurs at a larger $B_a$-value when $m$ is more negative.   Although the extreme limits are the same, again the intermediate behavior is different for the three systems.  As one would expect, the magnetic moment in the case of the model with overshoot reaches its $B_a \rightarrow \infty$ limit at a smaller applied field, than in the case without overshoot.  For small applied magnetic field we also observe oscillations in $M_{n,m}$ in case of a magnetic overshoot.  The oscillatory nature smoothes out and disappears as $B_a$ is raised or when the magnetic field is smoother at the edge (see Fig.~\ref{fig:moment}(c)).  This behavior can be understood by the following picture: depending on the radial quantum number $n$ the radial electron density has $n+1$ maxima.  For small magnetic fields the electron wavefunction is extended outside the magnetic anti-dot.  With increasing magnetic field the position of the maxima and minima in the electron density shift, and when a maximum is at the position of the overshoot, the magnetic field has the largest influence, and consequently the magnetic moment exhibits a minimum.  For a radial quantum number $n$, there will be $n$ maxima in the electron density which will shift through the overshoot at $\rho/a=1$ with increasing magnetic field strength $B_a$ and consequently $M_{n,m}$ exhibits $n$-maxima (one is at $B_a$=0).
When a minimum of the electron density is located at the overshoot, the magnetic moment has a local maximum.

In the superconducting disk case the limit $M_{n,m}(B_a\rightarrow\infty)=m$ is never reached.  In fact, $M_{n,m}$ slightly increases for $B_a/B_0>35$, and we found $M_{n,m}>m$ for all values of $B_a$.  This is a consequence of the small magnetic field under the disk, which is always present as one can see in Fig.~\ref{fig:profile}.  With increasing applied magnetic field $B_a$, this field grows continuously, which influences the magnetic moment.  The oscillatory nature of the magnetic moment of the superconducting disk case is still vaguely visible in the low magnetic field region.   

\section{Optical properties}
For the present case of magnetically confined dots the confinement potential is not quadratic like it is often assumed for the case of electrically confined dots.  This has important consequences for the optical absorption spectrum.  Due to the generalized  Kohn's theorem \cite{kohn} the long wavelength radiation couples only with the center of mass motion of the electrons in the quadratically confined dots and the absorption spectrum exhibits only two peaks.  In the present case we observe coupling between the center of mass motion and the other degrees of freedom.  Transitions are only possible for $\Delta m = \pm 1$ like in the case of quadratic confinement, but the other selection rule $\Delta n=0,1$ is now broken. This is shown in Fig.~\ref{fig:oscillatorprofile1} where the oscillator strength 
\begin{equation}
f_{0,0}^{n,m}=(2m_{e}/\hbar^2)(E_{n,m}-E_{0,0}){|\langle0,0|\rho e^{\pm i \varphi}|n,m\rangle|}^2,
\end{equation}
for excitation from the ground state is plotted as function of $B_a$.  This becomes more visible in Fig.~\ref{fig:oscillatorprofile2} which shows an enlargement of the smaller oscillator strength region.  

\begin{figure}
\caption{The transition energy for excitation from the ground state $(0,0)$ to $(n,m)$.  The solid curves correspond to the transitions with highest oscillator strength, the dashed and the dotted curves are respectively for transitions with amplitude one order and two orders of magnitude smaller. 
}
\label{fig:transitions}
\end{figure}
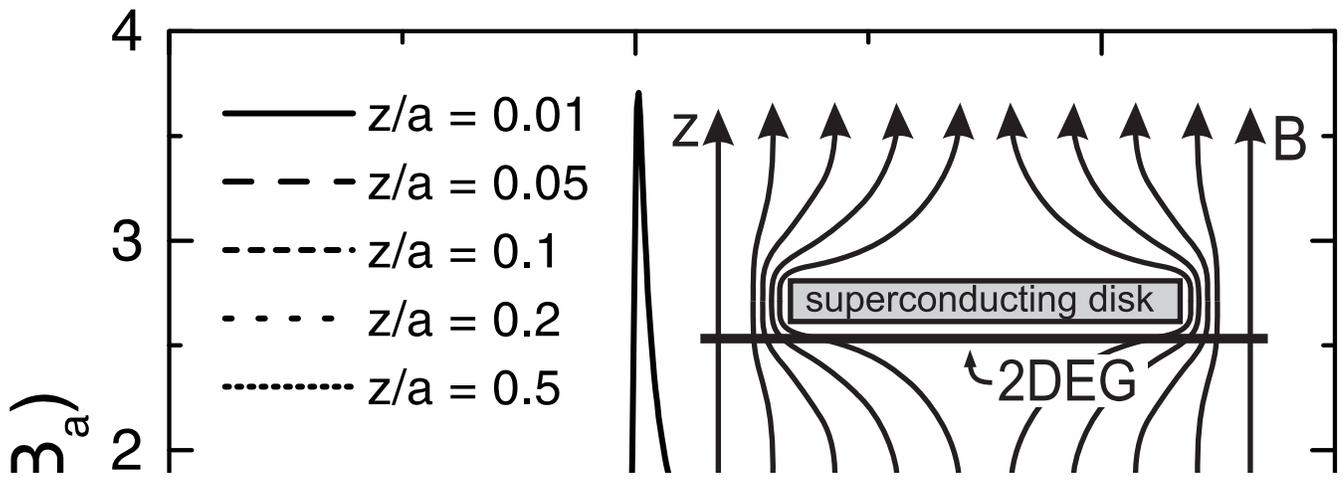
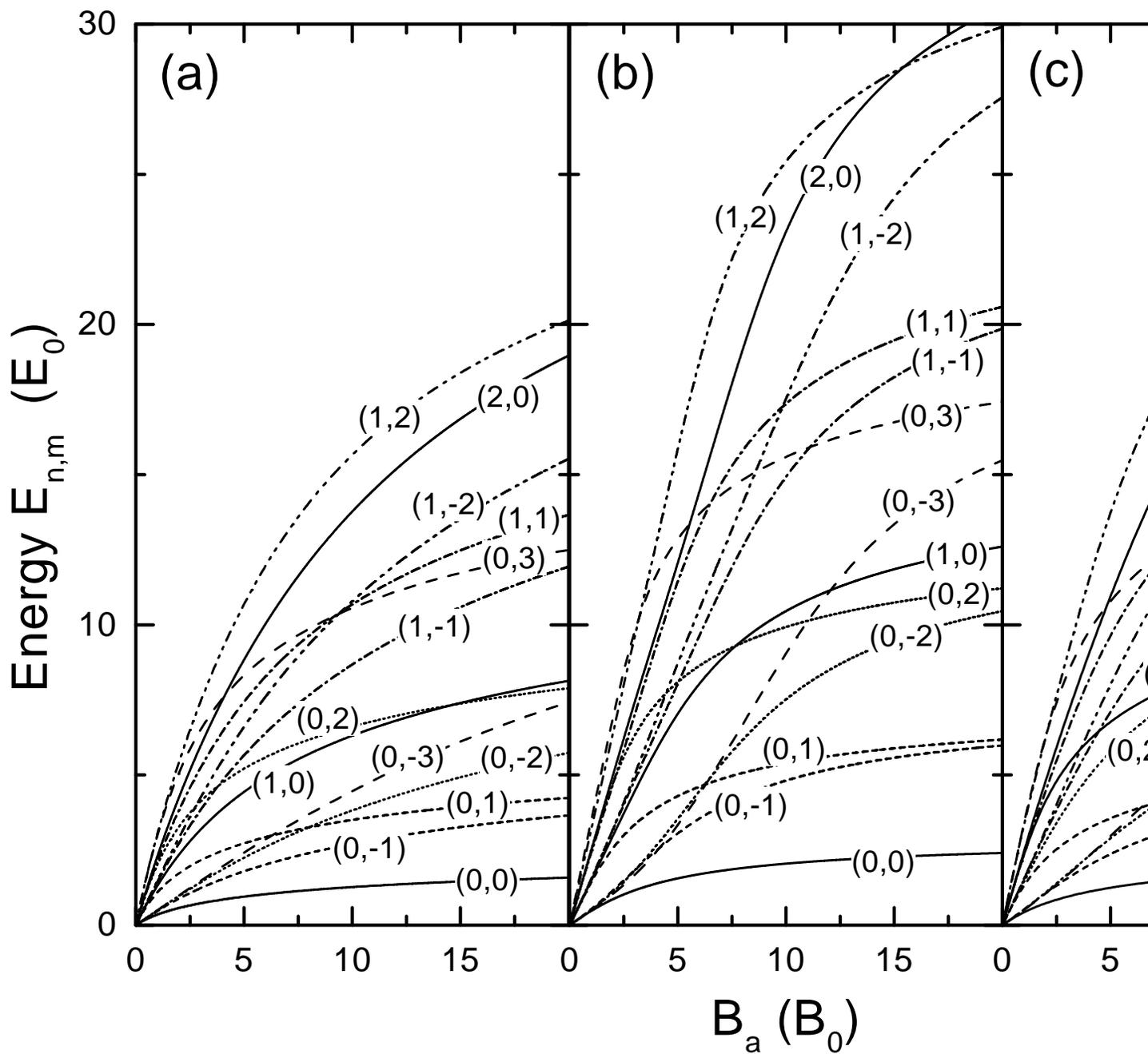
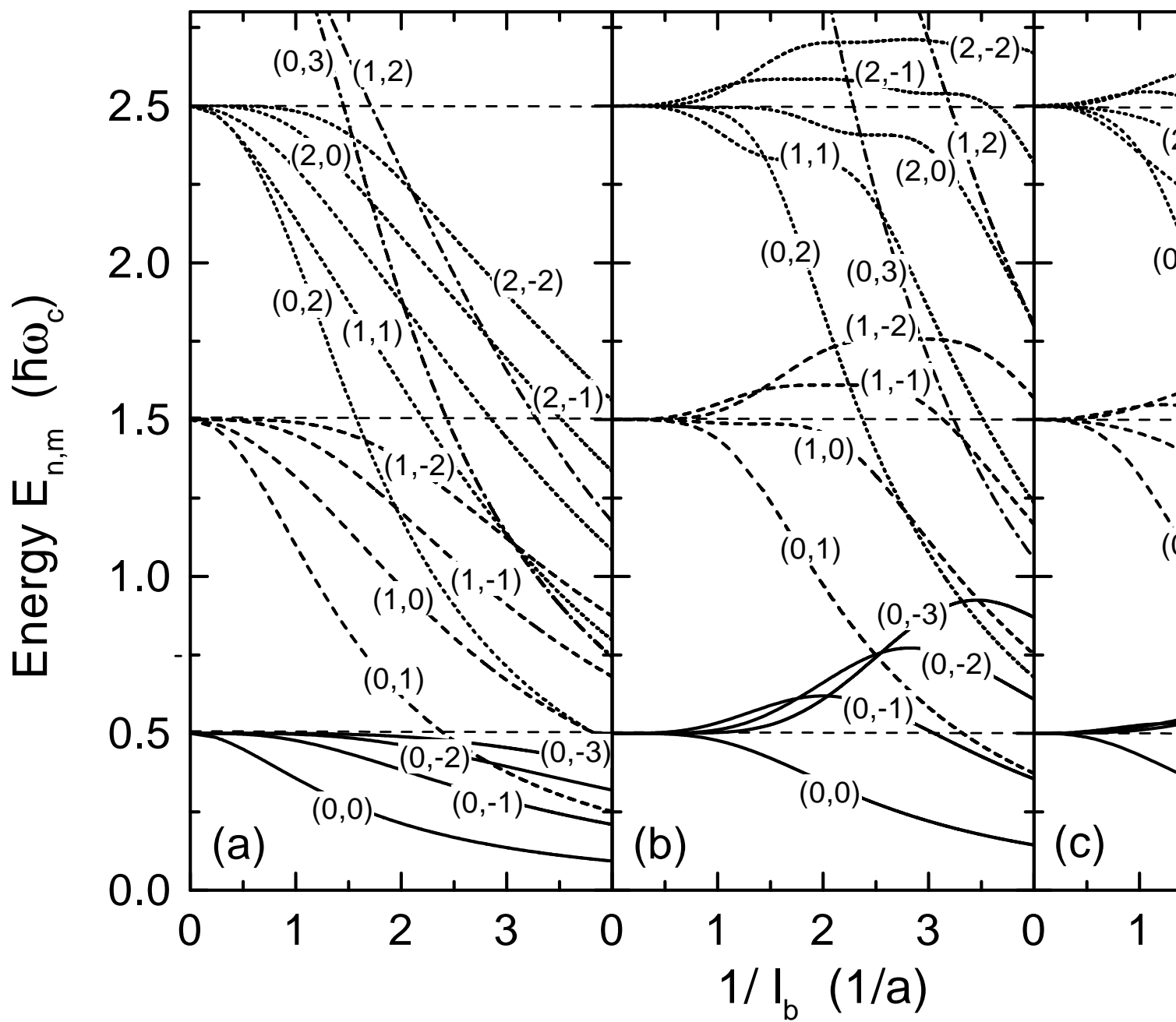
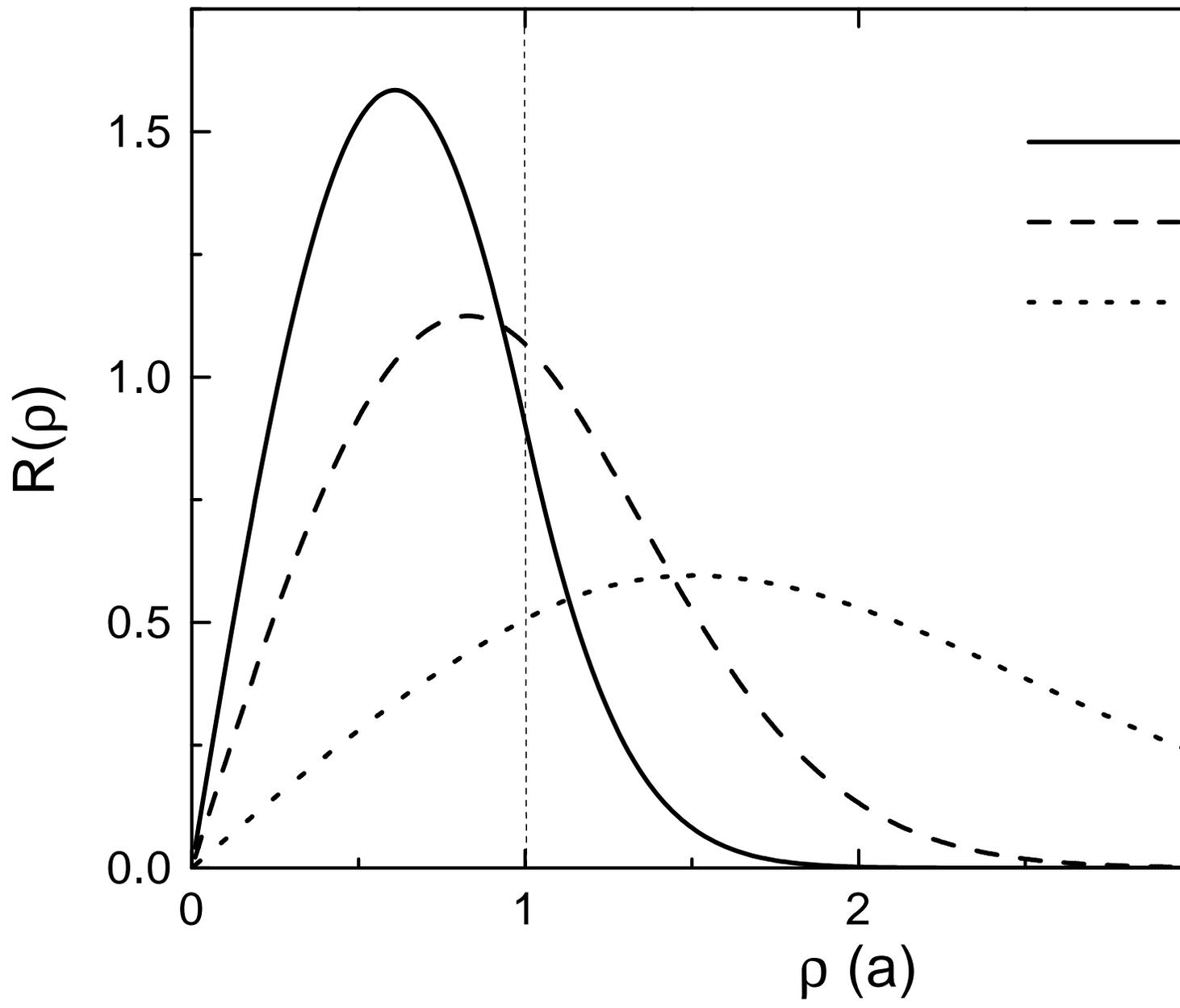
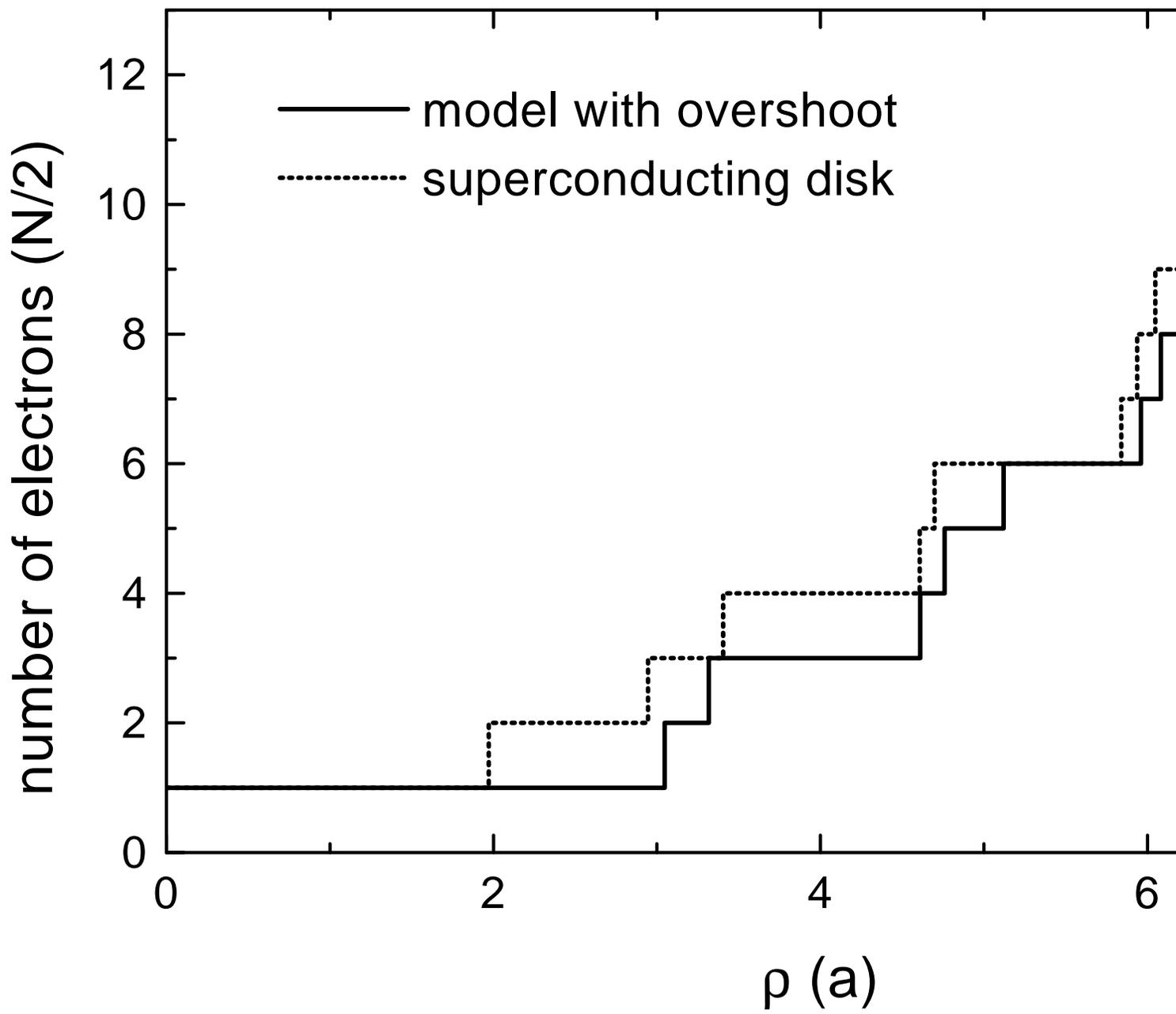
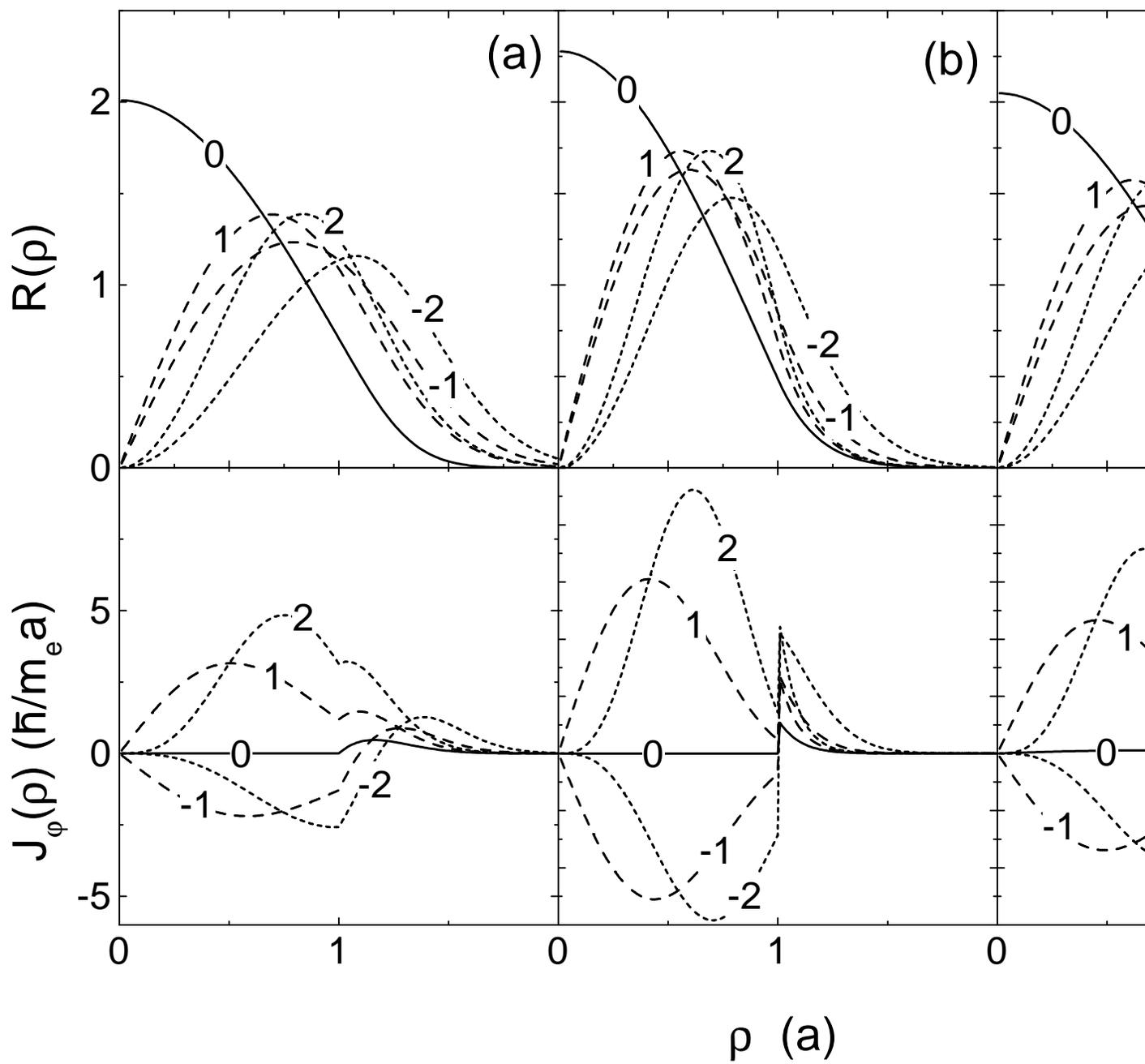
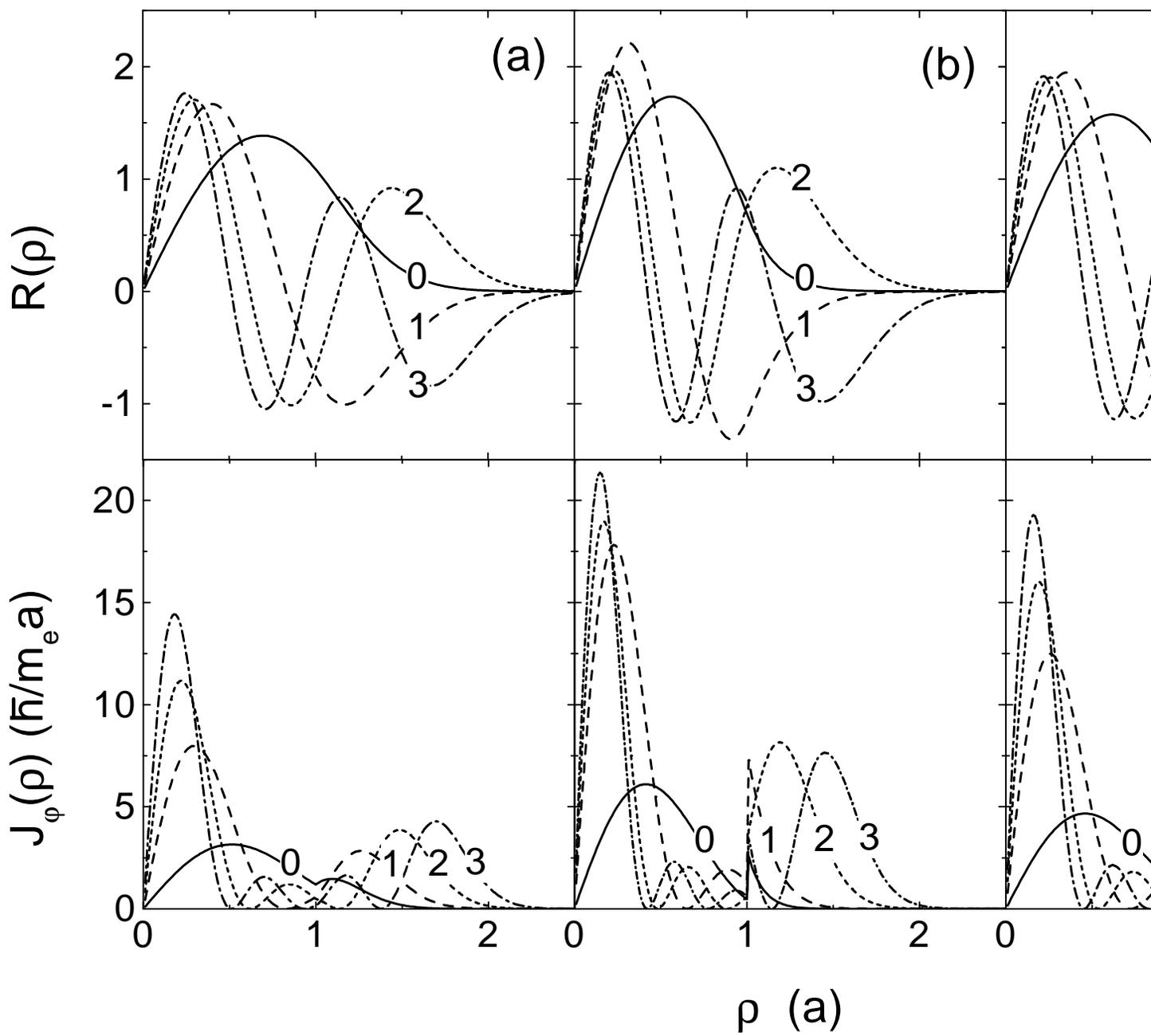
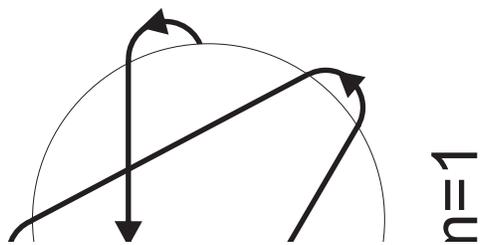
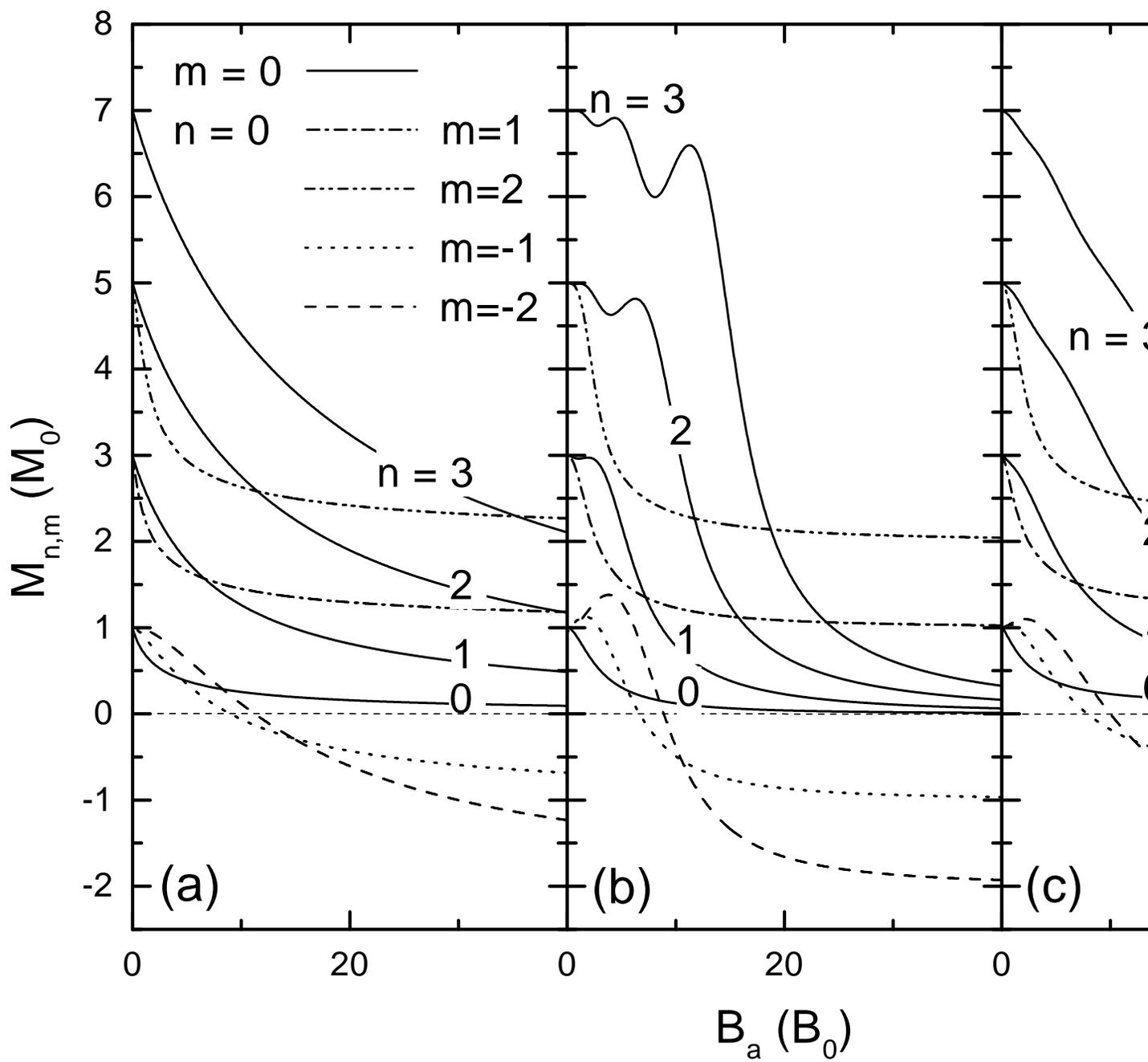
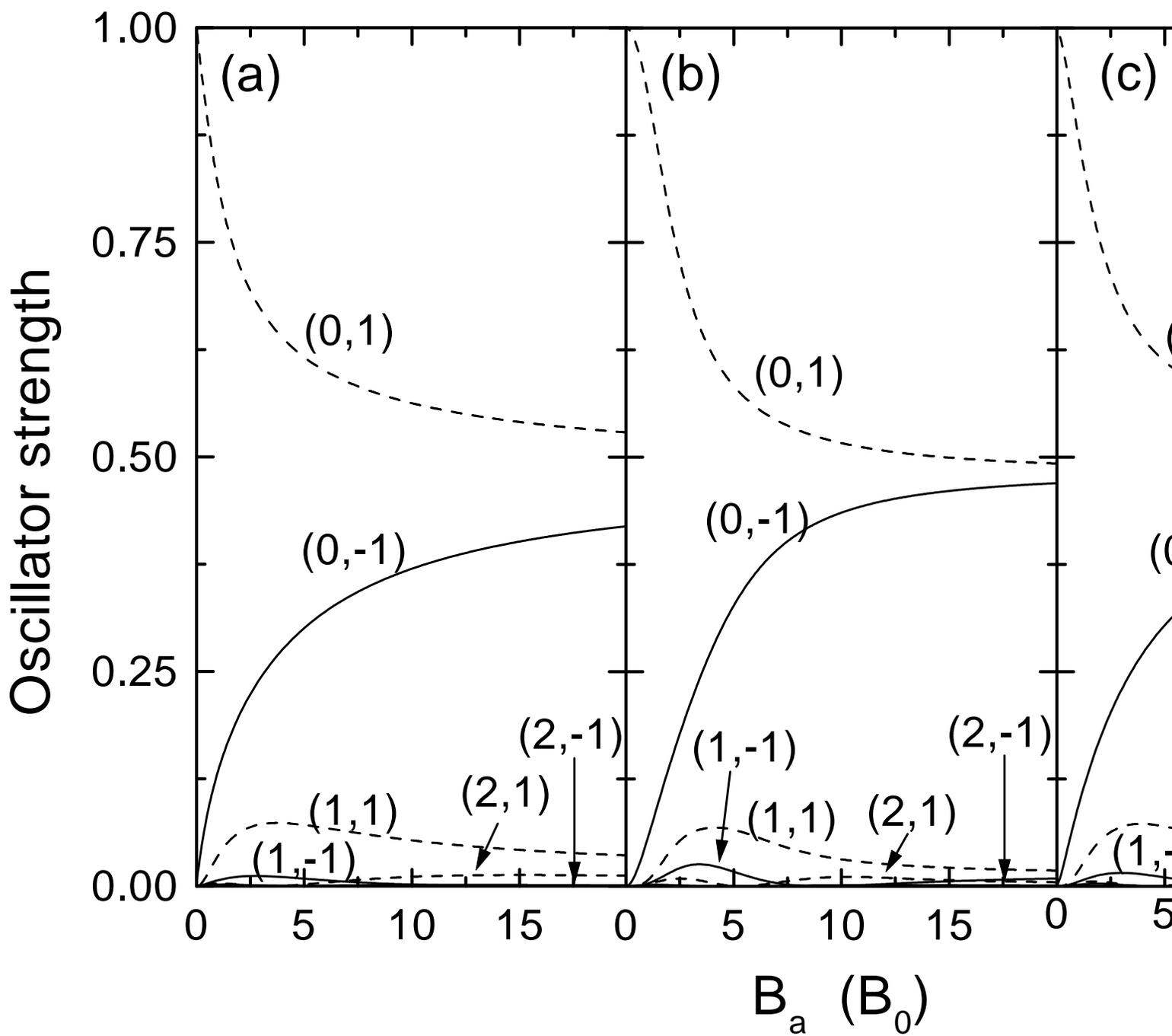
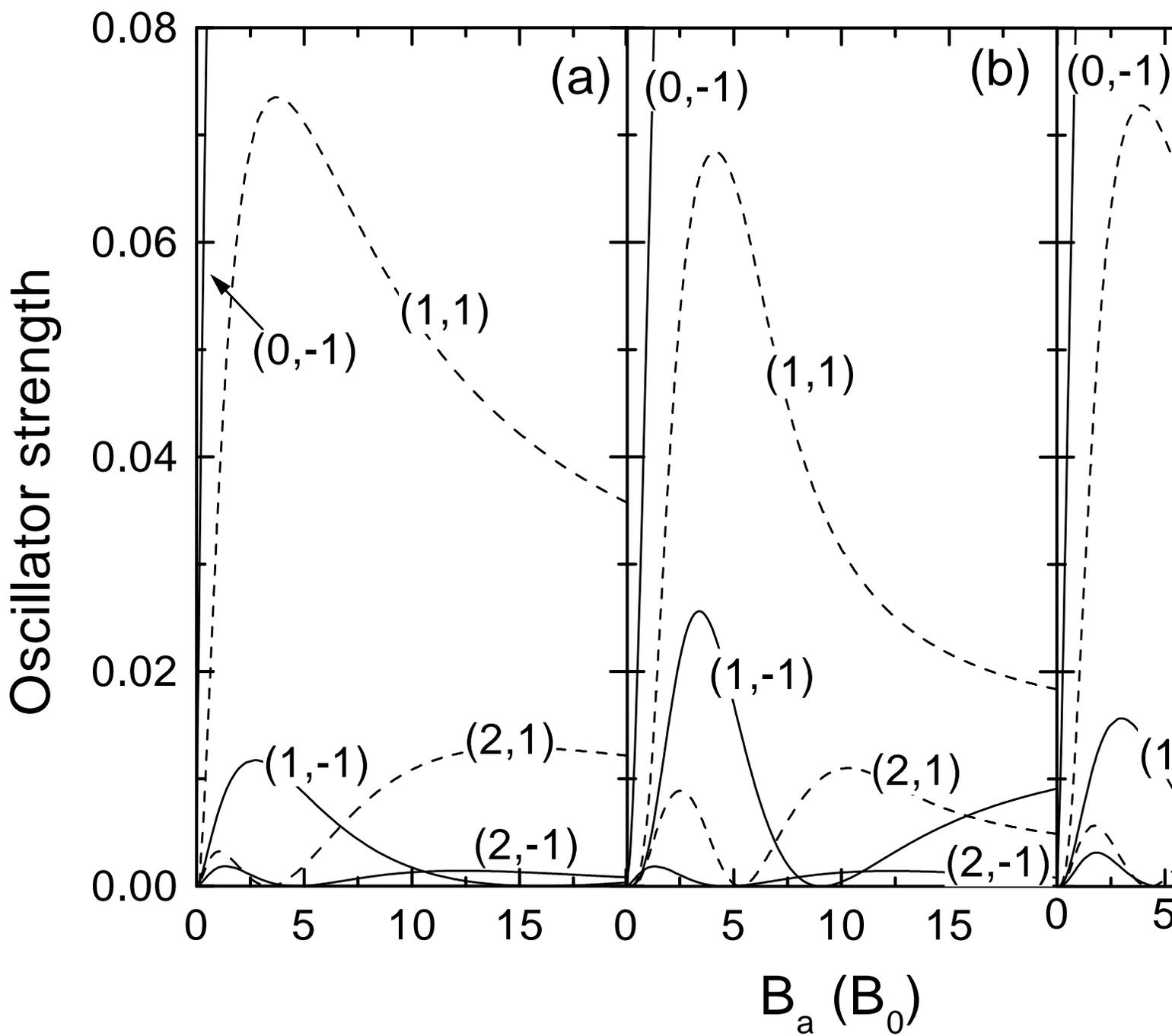
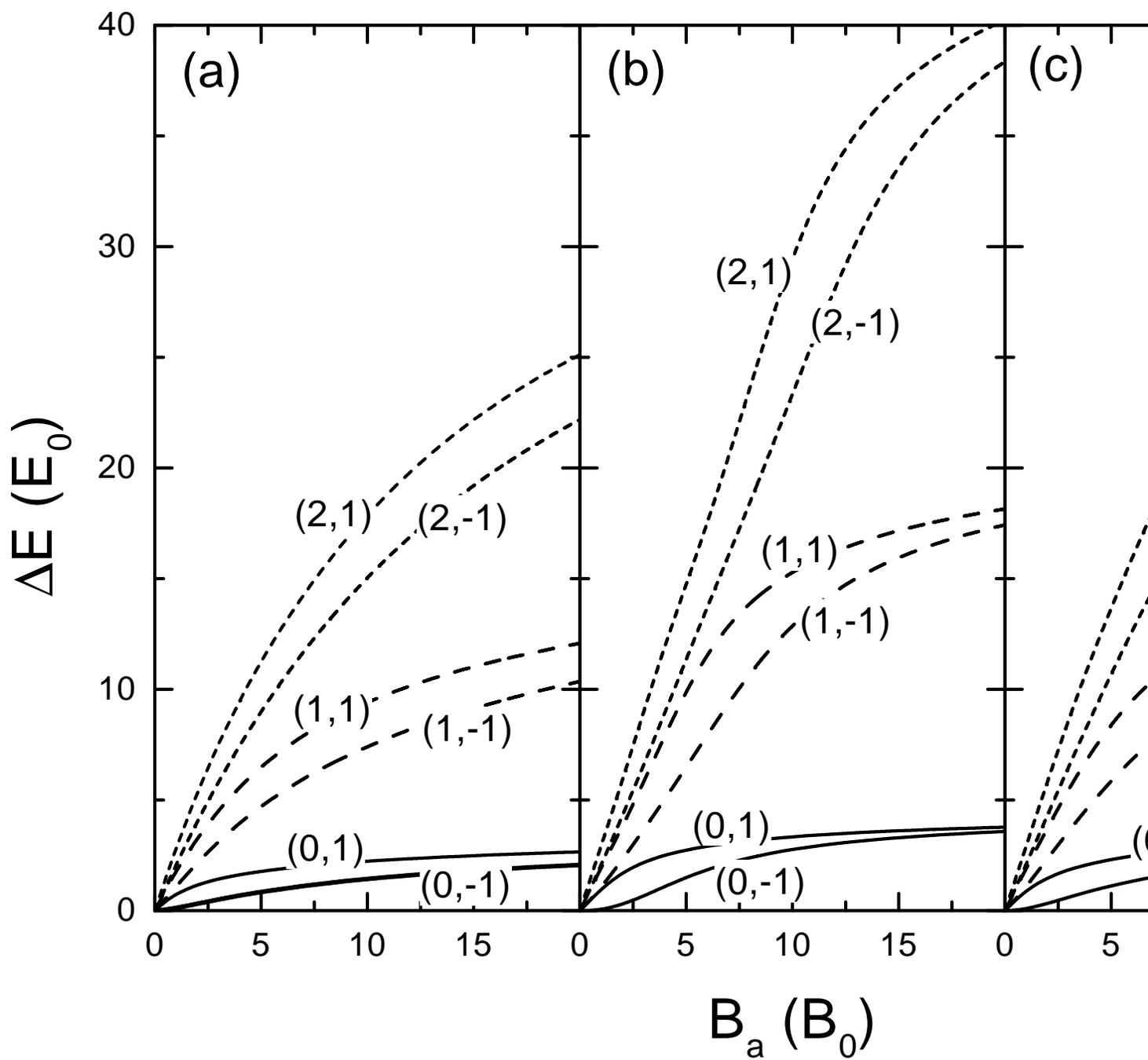

Consequently the absorption spectrum exhibits a larger number of peaks then in the case of a quadratic confinement potential.  The transition energies as a function of the applied magnetic field $B_a$ are plotted in Fig.~\ref{fig:transitions}.  The solid curves correspond to the transitions with largest oscillator strength, the dashed and the dotted curves are respectively for transitions with one order, and two orders of magnitude smaller oscillator strength.

\section{Conclusions}
In conclusion we have studied the single particle properties of a magnetically confined quantum dot.  When this is realized through a thin superconducting disk which is situated close to the 2DEG of a heterostructure, the magnetic field profile exhibits an overshoot at the edge of the disk which leads to a superlinear behavior of some of the energy levels as function of the strength of the external magnetic field $B_a$.  The consequence of this behavior is a discrete filling of the dot as function of $B_a$.  The circular current of the electron bound states has opposite sign inside and outside the dot region for the $m<0$ states.  These $m<0$ states exhibit a transition from a diamagnetic to a paramagnetic behavior with increasing $B_a$-field.  Because of the non quadratic nature of the confinement potential we predict that the optical spectra of these new dots exhibits a rich spectrum of lines.

\section{Acknowledgments}
This work was partially supported by a NATO-linkage Collaborative Research Grant, the Inter-university Micro-Electronics Center (IMEC, Leuven), the Flemish Science Foundation (FWO-Vl) and the IUAP-IV.  J. R. is supported by ``het Vlaams Instituut voor de bevordering van het Wetenschappelijk \& Technologisch Onderzoek in de Industrie'' (IWT) and F. M. P. is a research director with the FWO-Vl. 
%
%
%

%
%
%
%
\begin{figure}
\caption{(a) The calculated magnetic field profile in the 2DEG plotted as function of the radial distance from the center $\rho$ of the dot for different values of $z$, the distance of the disk above the 2DEG.  The inset is a side view of the experimental configuration we have in mind. (b) The vector potential profile in the 2DEG plotted as function of the radial distance from the center $\rho$ of the dot for the model systems without magnetic overshoot (dashed curve), with overshoot (dotted curve) and for the superconducting disk case with $z/a=0.01$ (solid curve) and $z/a=0.2$ (long dashed curve).}
\label{fig:profile}
\end{figure}
\begin{figure}
\caption{Energy spectra as function of $B_a$ for (a) the model without overshoot, (b) the model with overshoot, and c) the superconducting disk case with $z/a=0.01$.  The $B_a \rightarrow \infty$ limiting behavior is indicated at the right of the figure.   
}
\label{fig:spectra1}
\end{figure}
\begin{figure}
\caption{Energy spectra (in units of $\hbar \omega_c$) as function of $l_B/a \sim \sqrt{B_a}$ for (a) the model without overshoot, (b) the model with overshoot, and (c) the superconducting disk case with $z/a=0.01$.   
}
\label{fig:spectra2}
\end{figure}
\begin{figure}
\caption{Radial part $R(\rho)$ of the wavefunction with quantum numbers $(n,m)=(0,-1)$ in case of the model with overshoot as function of the radial distance to the center of the dot, for different values of $B_a$.
}
\label{fig:wavefunction}
\end{figure}
\begin{figure}
\caption{The filling of the magnetic anti-dot with electrons as function of $B_a$ for the model with overshoot (solid lines) and the superconducting disk model (dotted lines) with $z/a=0.01$.
}
\label{fig:filling}
\end{figure}  
\begin{figure}
\caption{The current density profile and the electron radial distribution when $B_a/B_0=10$ are shown for various $m$-values (indicated on the figure) with $n=0$ for (a) the model without overshoot, (b) with overshoot and (c) for the superconducting disk case with $z/a=0.01$.  
}
\label{fig:currentprofile1}
\end{figure}
\begin{figure}
\caption{The current density profile and the electron radial distribution when $B_a/B_0=10$ are shown for various $n$-values (indicated on the figure) with $m=1$ for (a) the model without overshoot, (b) with overshoot and (c) for the superconducting disk case with $z/a=0.01$.  
}
\label{fig:currentprofile2}
\end{figure}
\begin{figure}
\caption{Classical trajectories of electrons confined into a magnetic antidot without overshoot for $m=-1$ (clockwise), $m=0$ and $m=1$ (counterclockwise). 
}
\label{fig:trajectories}
\end{figure}
\begin{figure}
\caption{The magnetic moment $M_{nm}$ (in units of $M_0=-e\hbar/2m_e$) as a function of applied magnetic field strength $B_a$ for (a) the model without overshoot, (b) with overshoot and (c) for the superconducting disk case with $z/a=0.01$.
}
\label{fig:moment}
\end{figure}
\begin{figure}
\caption{The oscillator strength $f_{0,0}^{n,m}$ as a function of the applied magnetic field strength $B_a$ for (a) the model without overshoot, (b) with overshoot and (c) for the superconducting disk case with $z/a=0.01$. 
}
\label{fig:oscillatorprofile1}
\end{figure}
\begin{figure}
\caption{The same as Fig.~\ref{fig:oscillatorprofile1} but now the small oscillator strength region is enlarged.
}
\label{fig:oscillatorprofile2}
\end{figure}
\end{document}